\documentclass[useAMS,usenatbib]{mn2e}

\def\gsim{ \lower .75ex \hbox{$\sim$} \llap{\raise .27ex \hbox{$>$}} }
\def\lsim{ \lower .75ex \hbox{$\sim$} \llap{\raise .27ex \hbox{$<$}} }
\def\Mo{{\rm M_\odot}}

\title[Two body relaxation in CDM simulations]
{Two body relaxation in CDM simulations}

\author[J. Diemand et al.]
{Juerg Diemand,$$\thanks{diemand@physik.unizh.ch}
Ben Moore,$$ Joachim Stadel,$$ Stelios Kazantzidis,$$\\
$$Institute for Theoretical Physics, University of Z\"urich
,CH-8057 Z\"urich, Switzerland}

\begin{document}

\pagerange{\pageref{firstpage}--\pageref{lastpage}} \pubyear{2003}

\maketitle

\label{firstpage}

\begin{abstract}
N-body simulations of the hierarchical formation of cosmic structures
suffer from the problem that the first objects to form always contain
just a few particles. Although relaxation is not an issue for
virialised objects containing millions of particles, collisional
processes will always dominate within the first structures that
collapse.  First we quantify how the relaxation varies with
resolution, softening, and radius within isolated equilibrium and
non-equilibrium cuspy haloes.  We then attempt to determine how this
numerical effect propagates through a merging hierarchy by measuring
the local relaxation rates of each particle throughout the
hierarchical formation of a dark matter halo. The central few percent
of the final structures - a region which one might naively think is well 
resolved at the final time since the haloes contains $\approx 10^6$ particles 
- suffer from high degrees of relaxation.  It is not
clear how to interpret the effects of the accumulated relaxation rate,
but we argue that it describes a region within which one should be
careful about trusting the numerical results.
Substructure haloes are most affected by relaxation since they contain
few particles at a constant energy for the entire simulation.  We show
that relaxation will flatten a cusp in just a few mean relaxation
times of a halo. We explore the effect of resolution on the degree of
relaxation and we find that increasing $N$ slowly reduces the degree
of relaxation $\propto N^{-0.25}$ rather than proportional to $N$ as
expected from the collisionless Boltzmann equation. Simulated with the same
relative mass resolution (i.e. equal numbers of particles)
cluster mass objects suffer significantly more
relaxation than galaxy mass objects
since they form relatively late and therefore more of
the particles spend more time in small $N$ haloes.
\end{abstract}

\begin{keywords}
methods: N-body simulations -- methods: numerical --
dark matter --- galaxies: haloes
\end{keywords}

\section{Introduction}

A standard technique to study the formation and evolution of gravitating systems
is to perform an $N$-body simulation in which the mass distribution is discretised
into a series of softened point particles. This solution can be exact for a star
cluster where each particle represents a single star, but for cosmological simulations
of the dark matter each particle can be $10^{70}$ times larger than the 
GeV mass candidates
being simulated. In this approach the particles represent a coarse grained
sampling of phase space which sets a mass and spatial resolution. Unfortunately
these super-massive particles will undergo two body encounters that lead to energy
transfer as the system tends towards equipartition. In the real Universe the dark
matter particles are essentially collisionless and pass unperturbed past each other.

The processes of relaxation is difficult to quantify, but in the
large $N$ limit the discreteness effects inherent to the N-body technique
vanish, so one tries to use as large a number of particles as
computationally possible. Increasing the mass resolution of a given
simulation allows a convergence test of properties such as the dark
matter density profile i.e. \citet{Moore1998}, \citet{Ghigna},
\citet{Klypin} and \citet{Power}. These authors find that
to resolve the central one per cent of a dark matter halo the entire system 
must contain of the order a million particles. It is not known
what process sets this resolution scale since with one million
particles relaxation is expected to be small, even at one per cent of the virial
radius.

Unfortunately in most cosmological simulations
the importance of two body interactions does not vanish if
one increases $N$. Structure formation in the cold dark matter (CDM) model occurs
hierarchically since there is power on all scales, so the first objects
that form in a simulation always contain only a few particles 
\citep{Moore2001}, \citep{Binney2002}.
With higher resolution the first structures form
earlier and have higher physical densities because they condense out
of a denser environment. Two body relaxation increases with density,
so it is not clear if increasing the resolution can diminish the
overall amount of two body relaxation in a CDM simulation, i.e.  if
testing for convergence by increasing the mass resolution is
appropriate.

In isolated equilibrium systems relaxation rates can be measured from 
the energy dispersion. In cosmological simulations one can measure
the amount of mass segregation of multi-mass
simulations \citep{Binney2002} where lighter particles gain more
energy from collisional processes than the heavier particles. 
In Section \ref{estimate} we present a Fokker-Planck type relaxation time estimate, 
which was fitted to a series of test simulations (Section \ref{Equilibrium haloes}) where
we explore the relaxation rates as a function of N, radius and softening parameter
in both equilibrium and non-equilibrium cuspy haloes.
We then use this local relaxation rate estimate to follow the relaxation history
of each particle during 1000 time-steps of a hierarchical CDM simulation.
The resulting degree of relaxation as a function of spatial position within 
galaxy and cluster mass haloes is analysed in Section \ref{cosmo} and in 
Section \ref{discussion} we discuss the effects that relaxation 
has on haloes at $z=0$.

 \section{A local relaxation time estimate}\label{estimate}

In this paper we adopt the energy definition of the relaxation time \citep{Chandrasekhar}
stating that the mean relaxation time $T$ of a group of stars is the time
after which the mean square energy change due to successive encounters equals
the mean kinetic energy of the group: 
\begin{equation}  \label{T}
T = \Delta t \, \frac{\overline{E_{kin}}^2 }{\overline{\Delta E^2(\Delta t)}} \, ,
\end{equation}
where $\Delta E(\Delta t)$ is the energy difference of one particle
after time interval $\Delta t$ and the bar denotes the group average.

Note that this time is half of the relaxation time $T_v$ 
defined in \citet{Binney1987} who calculate
a mean velocity change,
because $\Delta E^2 / E^2 \simeq 2 \Delta v^2 / v^2$.
The orbit averaged Fokker-Planck estimate for $T_v$ is
\begin{equation}  \label{FP}
T_v = 0.34 \frac{\sigma^3}{G^2 m \rho \ln \Lambda} \, ,\; \Lambda \equiv \frac{b_{max}}{b_{min}}\, ,
\end{equation}
where $\sigma$ is the one dimensional velocity dispersion, 
$\rho$ the density and $m$ the 
particle mass. The parameters $b_{min}$ and $b_{max}$ are the minimum and maximum 
limits for the impact parameter.

To assess the degree of relaxation in cosmological simulations (section \ref{cosmo})
we estimate the local relaxation rate for each particle after every time-step and integrate
this up over the whole run:
\begin{equation}  \label{degree}
d(t_k) := \sum_{n=1}^{k} r_{LE}(t_n) \, \Delta t \, .
\end{equation}
For the local relaxation rate we use a formula similar to (\ref{FP})
\begin{equation}  \label{LE}
T_{LE}=\frac{1}{r_{LE}} = \gamma\frac{\sigma^3}{G^2m \rho C}\, ,\, \gamma \equiv 0.17 \,.
\end{equation}
The value of $\gamma$, and the parameters in the Coulomb logarithm $C$
are chosen to roughly fit measured relaxation times of equilibrium haloes, 
see section (\ref{spherical}). The Coulomb logarithm is
\begin{equation}  \label{Clocal}
  C \equiv 0.5\left[\ln(1 + \Lambda^2) - \frac{\Lambda^2}{ 1 + 
  \Lambda^2 } \right]
  \left( \simeq \ln \Lambda , \,\, \rm{if} \Lambda \gg 1\right) \, ,
\end{equation}
the analytical calculation for Newtonian potentials shows that 
$b_{min}=b_0=2Gm / v_{rel}^2$, $b_0$ is the impact
parameter where the deflection angle reaches $\pi/2$ \citep{Bertin}.
In a softened potential the scattering calculation has to be done numerically and
the results agree roughly with the Newtonian case if one sets $b_{min} = \epsilon$,
i.e. one ignores all encounters with an impact parameter smaller than the
softening length \citep{Theis}. We set
\begin{equation}  \label{bmin}
b_{min} \equiv \rm{max}(Gm/3\sigma^2, \epsilon) \simeq \rm{max}(b_0, \epsilon) \, ,
\end{equation}
because $\overline{v_{rel}^2} = 6\sigma^2$. The proper choice of $b_{max}$ is  
controversial, it is not clear whether it should be related to the size of the whole 
system or to the mean interparticle distance. For cosmological simulations we prefer
the second choice
\begin{equation}  \label{bmax}
b_{max} \equiv \beta (m/\rho)^{1/3} \, 
\end{equation}
because this is a local quantity that is easy to measure and less
ambiguous than defining the size of irregular shaped, collapsing structures.

We calculate the local velocity dispersion and density surrounding each particle
by averaging over its $16$ nearest neighbours. We do a simple top-hat average,
because using an SPH spline kernel leads to biased results
when using only $16$ particles. We found good agreement with all
measured relaxation rates (\ref{spherical}) when using  $\beta = 10$ and $\gamma = 0.17$. 
Averaging over different numbers of nearest neighbours 
the optimal parameters differ slightly due to different amounts of 
numerical noise in the local density and velocity dispersion.

\section{Two body relaxation in spherical halo models}\label{spherical}

In this section we present a number of test cases which we used to gauge 
our local estimate (\ref{degree}) for the degree of relaxation and show that it
agrees quite well (within 15 per cent) with
measured levels of relaxation for a wide range of particle numbers,
softening lengths and virial ratios. This range covers the haloes that
form in cosmological simulations, however all the test cases are
isotropic, spherical haloes. Haloes in cosmological simulations are
close to isotropic, but are triaxial and 
contain substructures. But one can argue that
locally the two are similar, and if a local relaxation time estimate
works in the spherical haloes, it should give a reasonable estimate
also in the cosmological case.

\subsection{Equilibrium haloes} \label{Equilibrium haloes}

In an equilibrium model the energy of each particle would be conserved 
in the $N \to \infty$ limit. For finite $N$ the energies of particles suffer
abrupt changes due to encounters.
Therefore we just have to measure these energy differences
to get the relaxation time with (\ref{T}).

Here we present a sequence of tests using spherical and isotropic 
Hernquist models \citep{Hernquist} which are a reasonable approximation
to the haloes found in cosmological N-body simulations.
(We found no difference between these simulations and
tests using \citet*{Navarro} and \cite{Moore1999}
profiles, constructed by solving for the exact phase space
distribution function numerically as described in \citet*{Kazantzidis}.)
The density profile of the Hernquist model is
\begin{equation}  \label{rhoHerquist}
\rho(r) = \frac{M}{2 \pi} \frac{a}{r} \frac{1}{(r+a)^3} \, .
\end{equation}
We set
\footnote{To rescale the results to different size haloes just change the distance scale 
by some factor $x \to fx$, mass scale $M \to f^3 M$ and the dynamical and relaxation
timescales do not change. To rescale to different timescales  $T \to cT$ 
do $M \to c^{-2} M$ with fixed length scale.}
the total mass to $M = 3.5\times 10^9 \Mo$ and the scale length to $a=10$ kpc.
Then the half mass radius is $r_h \simeq 2.4 a = 24$ kpc and the crossing time
at half mass radius is $T_{c} \equiv r_h / v_{circ}(r_h) \simeq 1.3$ Gyr.
 
All the simulations have been carried out using PKDGRAV, a state of the art, multi-stepping,
parallel tree-code \citep{Stadel}. The time-steps are chosen proportional to the square 
root of the softening length over the acceleration on each particle, 
$\Delta t_i = \eta\sqrt{\epsilon/a_i}$. We use $\eta = 0.25$, 
and a node-opening angle $\theta = 0.55$ for all runs in this section,
expect the long term integrations in subsection \ref{evolution}  
where we use $\eta = 0.03$.
Energy conservation was better than
0.1 per cent after several crossing times for all runs in this section. 

Due to softening the initial models are not exactly in equilibrium, 
the total kinetic energy is a few percent larger than half of the potential energy. For this
reason we evolved the models for five crossing times before measuring the energy dispersion,
which results in up to 10 per cent longer relaxation times.
 
\begin{figure}
\vskip 3.2 truein
\includegraphics{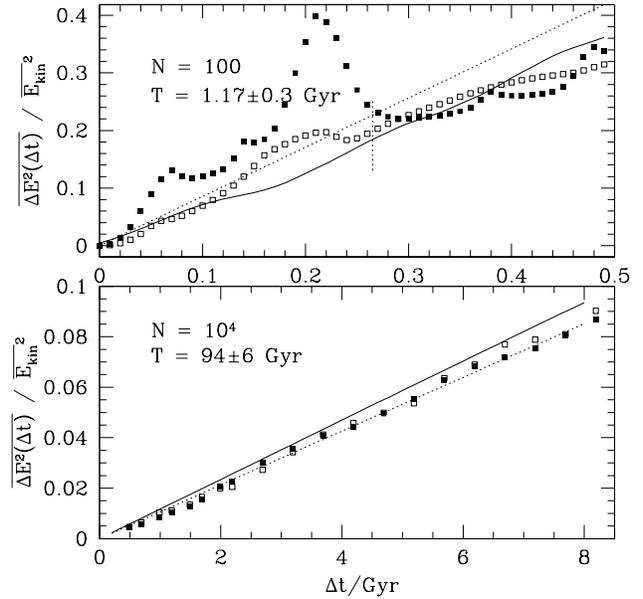}
\caption{\label{DeltaEvsT.ps}Mean squared energy change $\overline{\Delta E^2(\Delta t)}$ as a function of 
time. The filled squares show the mean squared energy change of the actual particles,
the open squares for the massless tracer particles. The dotted lines are linear fits to the 
mean squared energy change of the tracers, 
in the upper panel only the points left of the vertical bar are taken into account. 
The solid lines are averages of (\ref{degree}) over all massive particles.} 
\end{figure}

Figure \ref{DeltaEvsT.ps} shows $\overline{\Delta E^2(\Delta t)}$ as a function of 
time within haloes constructed with
$N=10^4$ and $N=100$ particles. The upper panel shows that for small 
numbers of particles $N \ll 10^3$ $\Delta E(\Delta t)$ becomes very noisy since 
there are fewer, but more significant encounters. To obtain more reliable results in small $N$ 
groups we added a sufficiently large number ($10^4$) 
of massless tracer particles following the same
distribution in real and velocity space as the massive particles. 
The open squares in Figure \ref{DeltaEvsT.ps} show 
the 'energy' dispersion of the tracers, which in large $N$ groups is just the same as the energy
dispersion of the massive particles, but it evolves much smoother with time in small $N$ groups.
We obtain the mean relaxation times (\ref{T}) of these haloes
with linear fits to the energy dispersion
of the tracer particles (open squares), taking into account points where
$\overline{\Delta E^2} < 0.2$, i.e. in the upper panel only the points
left of the vertical bar, to make sure that $\Delta t$ is small compared to the relaxation time.
The local relaxation estimate (solid line) (\ref{LE}) gives similar average degrees of relaxation in these
test cases (see also Figures \ref{TvsN.ps} and \ref{TvsEps.ps}). 

Note that the tracers are not in equilibrium with the halo, 
on average they gain speed in encounters and are ejected from
the core. Typically after one relaxation time the number of tracers inside of $r=a$
drops to one half of the initial number. 
Therefore it is important to use a $\Delta t$ shorter than $T$ and to add
the tracers after evolving the halo for five crossing times, otherwise
the relaxation in the core is not sufficiently reflected in the energy dispersion
of the tracers and $T$ could be underestimated significantly.

\subsubsection{Dependence on $N$ and $\epsilon$} \label{DepOnN}

\begin{figure}
\vskip 3.2 truein
\includegraphics{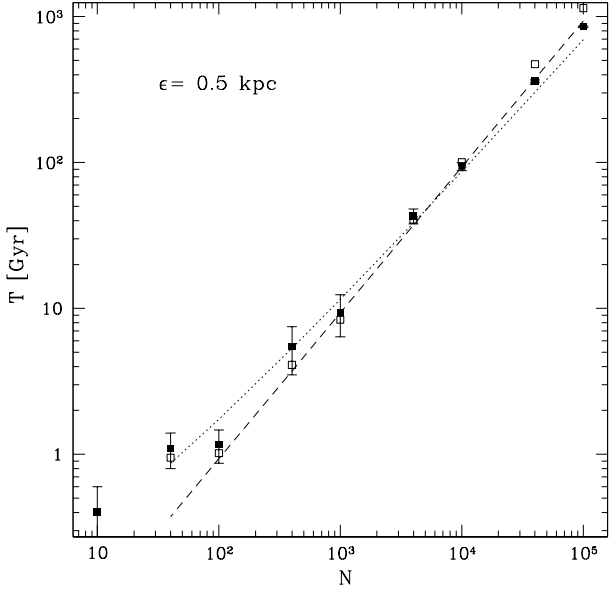}
\caption{\label{TvsN.ps}Average relaxation times of isotropic Hernquist models versus
particle number $N$, with a constant softening $\epsilon=0.5$ kpc.
The filled squares are the measured relaxation times, with error bars form the
linear fit of $\overline{\Delta E^2(\Delta t)}$. 
The open squares are the local estimates of the relaxation time (\ref{LE}).}
\end{figure}

\begin{figure}
\vskip 3.2 truein
\includegraphics{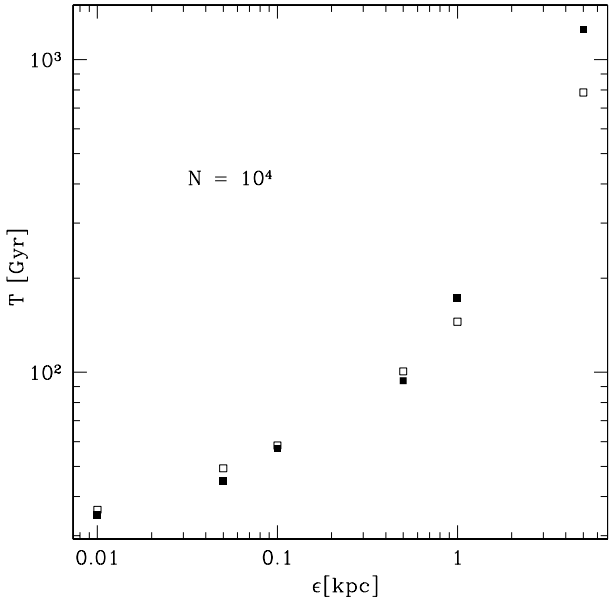}
\caption{\label{TvsEps.ps}Average relaxation times of an $N=10^4$ Hernquist models versus
softening length $\epsilon$. The filled squares are the measured relaxation times, 
the open squares are the local estimate (\ref{LE}).}
\end{figure}

Figures \ref{TvsN.ps} and \ref{TvsEps.ps} show the measured mean relaxation times
as a function of $N$ and softening parameter, $\epsilon$, 
compared with the average over all particles of the local relaxation estimate (\ref{LE}). 
We find that 
the measured relaxation times are proportional to $N$ (dashed reference line),
rather than to $N/\ln(N)$ (dotted line), as expected for a softened potential. 
The same result was found for King 
models by \citet*{Huang}. 
The local estimate of the relaxation time increases slightly faster with $N$ 
than the measured values. This is due to the fact that we 
choose a maximum impact parameter proportional
to the mean interparticle separation, i.e. $b_{max}\propto N^{-1/3}$, but the difference 
is less than 10 per cent for all relaxation times shorter than a Hubble time, i.e. for $N \le 10^4$. 

The dependence on the softening length is shown in Figure \ref{TvsEps.ps} for a
$N= 10^4$ model. The measured values (filled squares) increase 
faster with $\epsilon$ than the local estimate (open squares)
(like in Figure 2 of \citealt{Huang}), but the differences are small ($\gsim 15$ per cent) 
for realistic softenings $\epsilon \lsim 0.1 a = 1$ kpc.
The average relaxation time of this model increases from 30 Gyr to 180 Gyr when
the softening parameter is changed from 0.01 kpc to 1 kpc.
This is slightly higher than expected from the scaling 
with ${\rm ln}(\epsilon)$ since the density profile and central cusp 
are better resolved with smaller softening.

\subsubsection{Dependence on radius}

Measuring the relaxation time as a function of radius $T(r)$ proved to be quite
difficult.
The most credible method seems to take (\ref{T}) and replace the average
over all particles by the average over those which are in the corresponding radial bin at the 
beginning (or at the end) of the time interval $\Delta t$. Clearly one has to 
choose $\Delta t \ll \delta r/\sigma(r)$, where $\delta r$ is the size of the bins,
to make sure that most particles spend most
of $\Delta t$ in the same bin. This could also be achieved by placing the 
tracers on circular orbits, which leads to very similar results for $T(r)$ 
if $r \sim r_h$, but in to the centre this method fails, because there the 
circular velocity is much smaller than the velocity dispersion\footnote{From 
a convolution of the velocity distributions one finds 
that the relative velocities in encounters with tracers on circular orbits 
have a different dispersion ($\sigma^2_{v_{rel}} = 3\sigma^2 + v_{circ}^2$)
than those of encounters between the massive particles 
($\sigma^2_{v_{rel}}=6\sigma^2$). In non isothermal
haloes, $3\sigma^2 \neq v_{circ}^2$ and this leads to systematic errors in
the measured relaxation times. In Hernquist models $3\sigma^2 \sim v_{circ}^2$
holds only for $r\sim r_h$, and indeed we found good
agreement with the other methods only in this range.}.

\begin{figure}
\vskip 3.2 truein
\includegraphics{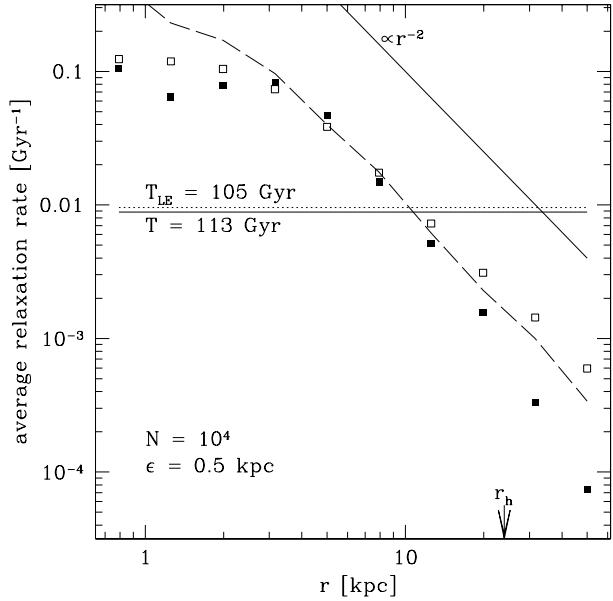}
\caption{\label{rel.hernquist.ps}Relaxation rate of an $N=10^4$ halo vs. radius.
The filled squares are the measured relaxation, 
the open squares are the local estimate $r_{LE}$, 
calculated from $16$ nearest neighbours during $0.1$ Gyr. 
The horizontal lines give the halo averages of measured (solid line) and 
estimated (dotted line) relaxation rates. We also plot $2/T_v$ (dashed line)
calculated from all particles in the radial bin.}
\end{figure}

In Figure (\ref{rel.hernquist.ps}) we plot the relaxation rate
against radius for a Hernquist model with $10^4$ particles.
We measured the energy dispersion (filled squares) during $\Delta t = 0.1$ Gyr for 
each particle, and averaged the values of particles starting in the same radial bin.
We also measured the local relaxation rate $r_{LE}$ (\ref{LE})
at $100$ time-steps during $\Delta t$ for each particle and summed them up. The radial 
averages are plotted with open squares. Again the agreement with the measured energy changes 
is better than 35 per cent except in the last three bins where the relaxation rates are
many thousands of Gyrs and the local relaxation measurement overestimates the true rate
of relaxation.  In the inner three bins the crossing times are shorter than $0.2$ Gyr, i.e. many of 
the particles had time to move through these bins during $\Delta t=0.1$ Gyr.

The dashed line is the inverse of half the Fokker-Planck estimate (\ref{FP}) 
with $\Lambda = r_h/\epsilon$, 
calculated using all particles in the bin, not only from $16$ nearest neighbours.
It scales like the phase space density $\rho(r)/\sigma^3(r)$ which 
scales almost exactly like $r^{-2}$ in a Hernquist model. 
The local estimate also follows this $r^{-2}$ scaling in the seven outer bins.
The measured relaxation scales more like $\propto r^{-3}$ in the outer region,
but the slope depends strongly on $\Delta t$, i.e. on how many particles from
the core with $100$ times higher relaxation rate have had time to reach the outer region.

The average relaxation times\footnote{Note that analytically
the average of the relaxation rate estimate $r_{LE}$ (\ref{FP}) is divergent for models 
with central cusps $\propto r^{-1}$ and steeper: $T = \int_0^R r_{LE}(r) \rho(r) r^2 dr 
\propto \int_0^R r^{-2} \rho(r) r^2 dr = \int_0^R \rho(r) dr$.} 
are about $10$ times shorter than
measured relaxation times at half mass radius, due to the fast relaxation in the 
high density core of cuspy haloes. For a less concentrated King model ($\Psi_0 = 5$)
the Fokker-Planck estimate seems to agree not only with the measured relaxation time 
at half mass radius, but also with the mean relaxation time \citep{Huang}.

\subsection{Non-equilibrium systems}\label{noneq}

The definition of the relaxation time (\ref{T}) is mostly used for systems
close to dynamical equilibrium like globular clusters, i.e. systems with constant
(or only slowly changing) mean kinetic energy. In this case the mean relaxation rate is 
also (roughly) constant and the degree of relaxation grows linear with time 
(like in Figure \ref{DeltaEvsT.ps}). 
Generalising this definition to non-equilibrium situations is
straightforward: The degree of relaxation of a group at some given time 
is the accumulated mean square energy change due to encounters
divided by the mean square kinetic energy at this time.

Now not all energy changes are due to encounters, so one needs another method to
measure the degree of relaxation. Again we use massless tracer particles like
in the last section. Instead of setting up their initial conditions exactly like those of
the massive particles, one can also restrict the tracers to a common 
orbital plane. In a spherical system this does not change their density profile nor the 
relative velocities in encounters with the massive particles.
If we choose the orbital plane of the tracers to be the $xy$-plane, 
then the accumulated energy change due to encounters $\Delta_{enc} E$ is
\begin{equation}  \label{vz}
\frac{ \overline{\Delta_{\rm{enc}} E(t)^2} }{ \overline{E_{kin}(t)}^2 } \simeq 
\frac{ \overline{\Delta_{\rm{enc}} E_{kin}(t)^2}  }{ \overline{E_{kin}(t)}^2 } \simeq
2  \frac{ \overline{\Delta_{\rm{enc}} v(t)^2}  }{ \overline{v(t)}^2 } \,
= 4 \frac{ \overline{v(t)_z^2}  }{ \overline{v(t)}^2 } \, ,
\end{equation}
as long as  $v_z (t)^2 \ll v(t)^2$, i.e. for small degrees of relaxation. 
This relates the energy dispersion to a more demonstrative
quantity and in the edge on view of the $xy$-plane one can actually observe 
the relaxation process since the degree of relaxation is roughly proportional 
to the thickness of the disk\footnote{Simulation movies are available at:
\mbox{www-theorie.physik.unizh.ch/$\sim$diemand/tbr/}}.
When the amount of relaxation approaches unity (\ref{vz})
tends to underestimate the degree of relaxation, because tracers on new
out of plane orbits will eventually reach a turnaround point where $v_z = 0$.
Also the probability that one tracer suffers more than one encounter grows with time,
therefore one should place a new set of tracers into the plane 
to get an accurate relaxation rate as soon as the 
amount of relaxation is close to unity.

With (\ref{vz}) we can measure the amount of relaxation in
non-equilibrium situations, we only need one symmetry plane to be 
able to apply this method, therefore it allows us to measure the amount of relaxation 
during a collapse or a merger.

\subsubsection{Collapsing haloes}\label{collapse}

In CDM simulations the virial ratio $\alpha \equiv 2 E_{kin}/ |E_{pot}|$ 
is close to zero at the beginning of a halo collapse
and grows towards unity as the halo reaches dynamical equilibrium. 
In the previous sections
we showed that the local estimate (\ref{LE}) works for $\alpha = 1$, 
but for $\alpha \to 0$ the phase space density $\propto \alpha ^{-1.5}$ goes to infinity. 
Down to which virial ratio can we trust our local estimate?

\begin{figure}
\vskip 3.2 truein
\includegraphics{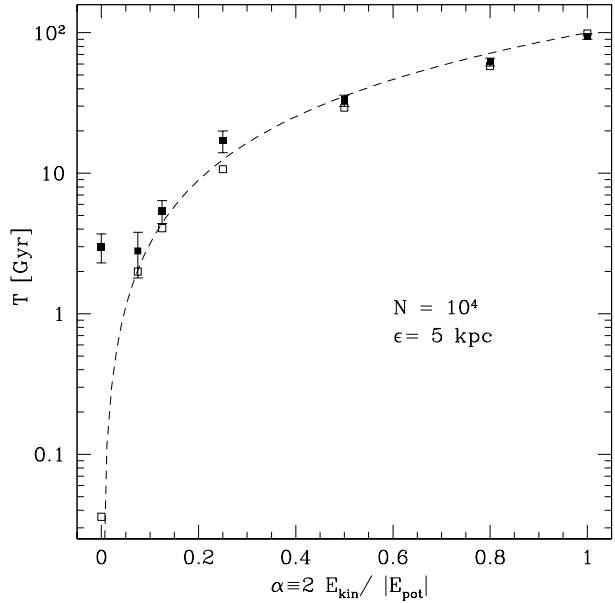}
\caption{\label{TvsA.ps}Relaxation times of a $N=10^4$ non-equilibrium 
Hernquist model vs. virial ratio. The filled squares are 
measured with $\overline{v_z^2} / \overline{v}^2$, 
the open squares are the local estimate (\ref{LE}). 
The reference line is $\propto \alpha^{1.5} \propto v^3/\rho$.}
\end{figure}

\begin{figure}
\vskip 3.2 truein
\includegraphics{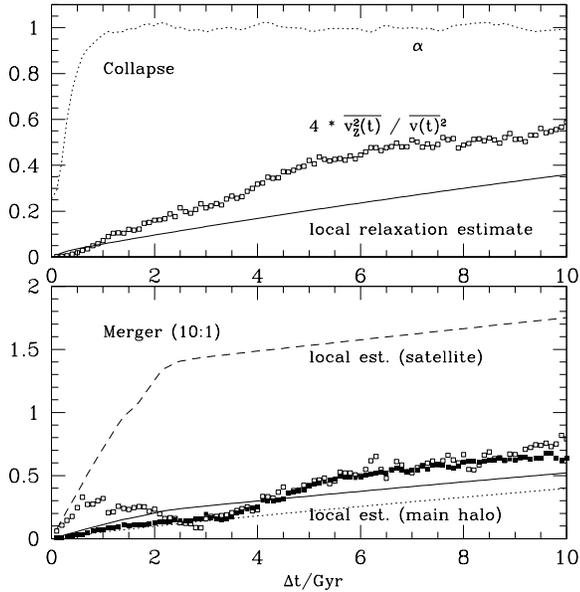}
\caption{\label{DeltaV.ps}Average degrees of relaxation during a collapse of an 
$N=10^4$ Hernquist halo, the initial virial ratio $\alpha$ is $0.25$ (top).
The lines are the average of the local relaxation estimate (\ref{LE}), the squares are the
amount of relaxation measured from a group of tracer set in a plane initially (\ref{vz}).
The bottom panel shows average degrees of relaxation during a a merger of 
an $N=10^3$ halo into a ten times more massive system with $N=10^4$. 
Here the open squares are the measured average for the satellite and the filled squares
for the main halo.}
\end{figure}

To answer this question we began with equilibrium Hernquist models 
(same parameters as in section \ref{DepOnN}) with $10^4$ particles
and multiplied all velocities with $\sqrt{\alpha}$. Therefore $\sigma^3 \propto \alpha^{1.5}$
and the phase space density $\rho/\sigma^3 \propto \alpha^{-1.5}$. To these models 
we added  $10^4$ massless tracer particles with the same phase-space distribution and
tilted their orbital planes into the $xy$-plane. Then we measure
$\overline{v_z^2} / \overline{v}^2$ and the local estimate (\ref{LE}) 
at $100$ time-steps during the first $0.1$ Gyr of the collapse. Linear fits give the 
relaxation times plotted in Figure (\ref{TvsA.ps}), the dashed line shows the
scaling $\propto \alpha^{1.5}$ expected from the Fokker-Planck type estimates 
(\ref{FP}) and  (\ref{LE}), since these times are proportional to 
one over phase-space density. Our relaxation time estimate becomes very small 
when the virial ratio goes to zero, but the measured relaxation times remain
on the order of a few dynamical times. $T_{LE}$ is within a factor of two 
for $\alpha \gsim 0.075$ and within 10 per cent for $\alpha \gsim 0.5$.
Also after the first $0.1$ Gyr the local estimate
follows the measured values, one example ($\alpha=0.25$) 
is plotted in the top panel of Figure (\ref{DeltaV.ps}).

\subsubsection{Mergers}\label{mergers} 

The last test case for the local relaxation estimate is a merger of a small system into 
a more massive one. The problem in this case is that the small halo gains a lot of kinetic
energy when falling into the main halo, so its accumulated energy changes due to encounters
can become smaller {\it relative} to the mean kinetic energy of the group. A local estimate
can never capture this decrease since it can not know about the gain in external kinetic energy.
For a surviving subhalo one can argue that its accumulated energy changes should still 
be compared to its roughly constant internal kinetic energy
to get an estimate of how affected it is by relaxation.
But for haloes that are disrupted (and stripped particles from subhaloes) one has to worry
about how their overestimated degrees of relaxation affect the average relaxation of
the main halo.

The bottom panel of Figure (\ref{DeltaV.ps}) shows how relaxation develops in a head on merger
of a $N=10^3$, $a=3$ kpc halo into a ten times more massive halo with $N=10^4$ and $a=10$ kpc.
The initial separation was 100 kpc with a small relative velocity of $2.2$ km/s. The satellite
falls in and reaches the centre of the main halo after 3 Gyr. Note how the relaxation of
the satellite (open squares) {\it decreases} during infall, this can not be followed by the average
local estimate (\ref{LE}) of the satellite (dashed line), but still the average over the whole
system (solid line) gives a good estimate for the mean degree of relaxation. We also verified this
for equal mass mergers, there the decrease during infall is small because both haloes 
have quite large internal energies initially.

\subsection{Evolution of isolated Halos}\label{evolution}

The dynamical evolution of globular clusters is 
driven by relaxation, which can lead to core collapse and
evaporation on a timescale of a few tens of half mass relaxation
times, i.e. the core loses energy to an expanding 
outer envelope of stars 
and gets denser and hotter (\citealt{Binney1987}; \citealt{Spitzer}).

In the next section we show that haloes in cosmological simulations
are typically between one and ten mean relaxation times old. (In terms 
of the much longer half mass relaxation time they are younger than 
one or two half mass relaxation times.) Therefore we do not expect that 
density profiles in cosmological simulations are significantly affected by
the core collapse process.

\begin{figure}
\vskip 3.2 truein
\includegraphics{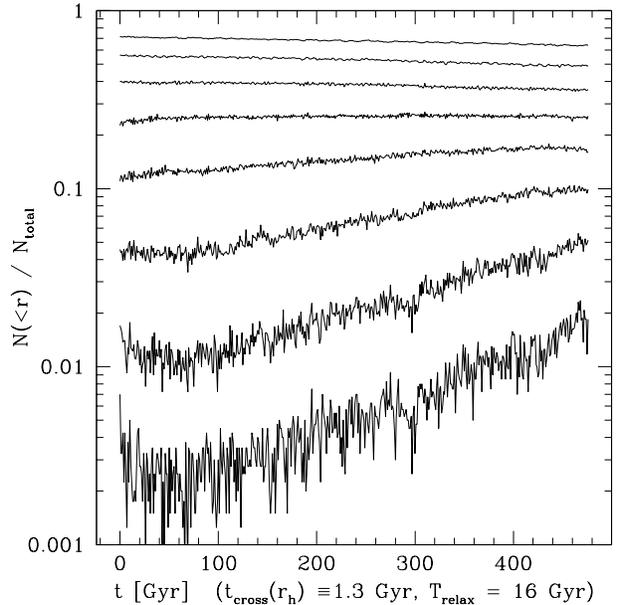}
\caption{\label{NvsT.long.ps}From top to bottom, the mass fraction within
$0.89, 1.58, 2.81, 5.00, 8.90, 15.8, 28.1$ and $50.0$ kpc 
of a $N=4'000$ Hernquist models vs. time.}
\end{figure}

Studies of globular cluster evolution start with models that are isothermal in
the centre (e.g. Plummer spheres, King models) and then show a slow but monotonic
density increase in the core. In contrast the cuspy haloes in cosmological simulations are
not isothermal: the velocity dispersion decreases in the central regions.
In this case relaxation
leads to an energy flow inwards, the core evolves towards a less dense, isothermal 
state first. Later the system evolves just like the models in globular cluster
calculations \citep{Quinlan}. The N-body simulations of \citet{Hayashi} show this evolution 
starting from an NFW profile. We confirmed their result by evolving an
$N=4'000$ Hernquist model for $360$ crossing times 
(see Figure \ref{NvsT.long.ps}). 

\begin{figure}
\vskip 3.2 truein
\includegraphics{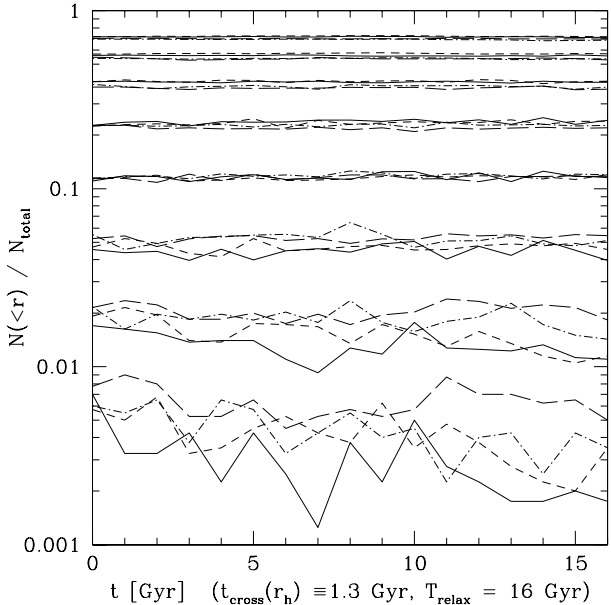}
\caption{\label{NvsT.ps}From top to bottom, the mass fraction within
$0.89, 1.58, 2.81, 5.00, 8.90, 15.8, 28.1$ and $50.0$ kpc 
of five $N=4'000$ Hernquist models vs. time.}
\end{figure}

Figure \ref{NvsT.ps} shows the evolution 
of five $N=4'000$ Hernquist models during $16$ Gyr, using
a softening of $\epsilon = 0.1 $kpc. For this long term evolution we 
use a more conservative time step parameter $\eta=0.03$, energy is conserved
within 0.36 per cent even after $360$ crossing times. The crossing time
at the half mass radius is $1.3$ Gyr, the initial mean relaxation time is $16$ Gyr
and the initial half mass relaxation time is $71$ Gyr. After $50$ Gyr this halo
is about three mean relaxation times old, a realistic value for haloes in current 
cosmological simulations, see section (\ref{cosmo}). The same would happen
to a $N=100$ halo in only $1.25$ Gyr, we use $N=4'000$ just to have a well defined 
density profile down to $0.1 a = 1$ kpc.
 
\begin{figure}
\vskip 3.2 truein
\includegraphics{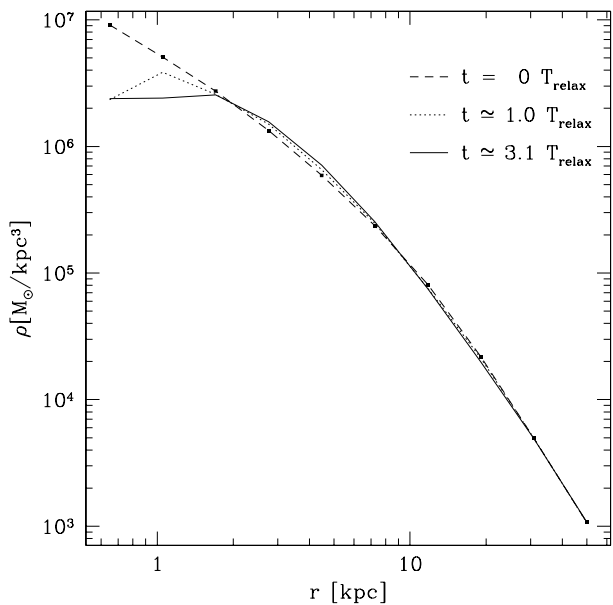}
\caption{\label{hernquist.pro.ps}
Averaged density profiles of five $N=4'000$ Hernquist models
(same models as in Figure \ref{NvsT.ps}),
initial profile (dashed line), after $16$ Gyr (dotted line)
and after $50$ Gyr (solid line).
The points indicate the radius of the outer borders of the spherical bins.}
\end{figure}

Within few mean relaxation times the velocity dispersion rises in the centre and 
at the scale radius it drops slightly, in the end the core is
already close to isothermal. This energy gain is compensated with core expansion, the inner two 
mass shells plotted in Figure \ref{NvsT.ps} clearly lose mass. This is not a 
numerical artifact, the softening we used is much smaller than the inner bin
and in a $N=4\times 10^4$ reference model the mass loss in the inner two bins
is about ten times slower, i.e. this
is really an effect driven by relaxation. The corresponding density profiles 
are less steep in the inner 3 per cent of the halo, see Figure \ref{hernquist.pro.ps},
and show constant density cores in this region.

\section{Relaxation in Cosmological Simulations} \label{cosmo}

Here we present results from four low to medium resolution 
$\Lambda$CDM simulations ($\Omega_{\Lambda} = 0.7$, $\Omega_m = 0.3$,
$\sigma_8 = 1.0$). We generate initial conditions with
the GRAFIC2 package \citep{Bertschinger}.
We start with a $128^3$ particle cubic grid with
a comoving cube size of $60 $Mpc (particle mass $m_p = 3.6\times 10^9 \Mo$).
Later we refined two interesting regions, in the first one 
a cluster halo ($M_{200} = 7.4\times 10^{13} \Mo$, 
$r_{200} = 1440$ kpc) forms, the refinement
factors are 2 and 3 in length, ie. 8 and 27 in mass (run C2,C3).
In the second region a galaxy size halo 
($M_{200} = 1.4 \times 10^{12} \Mo$, $r_{200} = 350$ kpc)
forms. There we used
a refinement of 9 in length, ie. 729 in mass, and included a buffer
region, about $2$ Mpc deep, with an intermediate refinement factor of 3
(run G9). We start the simulations when the standard deviation of the density 
fluctuations in the refined region reaches $0.2$. 
The softenings used in the refined regions are
$\epsilon=1.86$ kpc for the cluster and $\epsilon=0.5$ kpc for the galaxy. 
i.e. $\epsilon \simeq 0.0013 r_{vir}$ in both cases. We also run the 
unrefined cube again with this softening in the cluster forming region (run C0).
The numerical parameters are as in section \ref{Equilibrium haloes},
but at late epochs we use a larger node-opening angle to speedup
the runs, $\theta = 0.7$ for $z < 2$. 

In non equilibrium, non spherical haloes of a cosmological simulation it 
is not possible to measure two body relaxation times with the methods used in 
section \ref{spherical}. 
\citet{Binney2002} used initial conditions 
with two species of particles with different mass.
Both species start from a regular lattice, such that the nodes of one
grid are at the centres of the cells of the other, and both are then displaced according to 
the Zel'dovich approximation. In a collisionless simulation the final
distribution of the particles would be independent of mass. They
found differences in the number density of the light and
heavy particles in the centres of haloes. 

Here we compliment the study of Binney \& Knebe by applying the results of the previous
sections to cosmological simulations.
We assign a degree of relaxation $d$ to each particle, which is calculated after
each of $1000$ fixed time-steps $\Delta t$ from the local relaxation rate estimate (\ref{LE}) 
and summed up over the whole cosmological simulation (\ref{degree}).

As shown in section \ref{collapse}, $r_{LE}$ reflects the measured relaxation rates only
for virial ratios $\alpha \gsim 0.1$. Since this is not case for the first steps in a CDM simulation,
we set $r_{LE}$ to zero before the local density reaches some threshold. When the local
density reaches $6 \rho_0$ ($\simeq$ density at turnaround in the spherical collapse) the typical 
values for $\alpha$ are close to $0.4$, later at $170 \rho_0$ ($\simeq$ density at 
virialisation in the spherical collapse) $\alpha$ is close to unity. We used these two
density thresholds, in the first case we write $d_{\rm{TA}}$ for the ``degree of relaxation
since turnaround'', otherwise $d_{\rm{VIR}}$ for ``degree of relaxation
since virialisation''. The relaxation averages over all particles inside $r_{200}$ 
and $0.1 \,r_{200}$ at $z=0$ are given in table \ref{tab}.

\begin{table}
\caption{Average Degrees of Relaxation}
\label{tab}
\begin{tabular}{l || c | c | c || c}
               Run & C0 &  C2 & C3  & G9\\
      \hline
      \hline
      $N_{200}$ & $20'500$ & $177'000$& $650'000$ & $250'000$\\
      \hline
      \hline
      $d_0$ inside $0.1 r_{200}$ & $5.98$ & $3.62$ & $3.06$ & $1.67$ \\
      $d_0$ inside $r_{200}$ & $5.23$ & $3.34$ & $2.52$ & $1.15$ \\
      \hline
      $d_{\rm{TA}}$ inside $0.1 r_{200}$ & $4.74$ & $2.40$ & $1.78$ & $0.72$ \\
      $d_{\rm{TA}}$ inside $r_{200}$ & $3.67$ & $2.42$ & $2.12$ & $1.02$ \\
      \hline
      $d_{\rm{VIR}}$ inside $0.1 r_{200}$ & $3.58$ & $1.61$ & $1.17$ & $0.42$ \\
      $d_{\rm{VIR}}$ inside $r_{200}$ & $2.50$ & $1.52$ & $1.34$ & $0.58$ \\
\end{tabular}
\end{table}

\subsection{Number of particles}

A reassuring result is that with more particles the simulations are 
less affected by two-body relaxation, even though one resolves more 
small $N$ progenitors. This confirms the significance of convergence tests
that vary the number of particles. The average degree of relaxation inside of
$0.1 r_{200}$ scales like $N^{-0.3}$, and the relaxation inside of
$r_{200}$ like $N^{-0.2}$.

\begin{figure}
\vskip 6.6 truein
\includegraphics{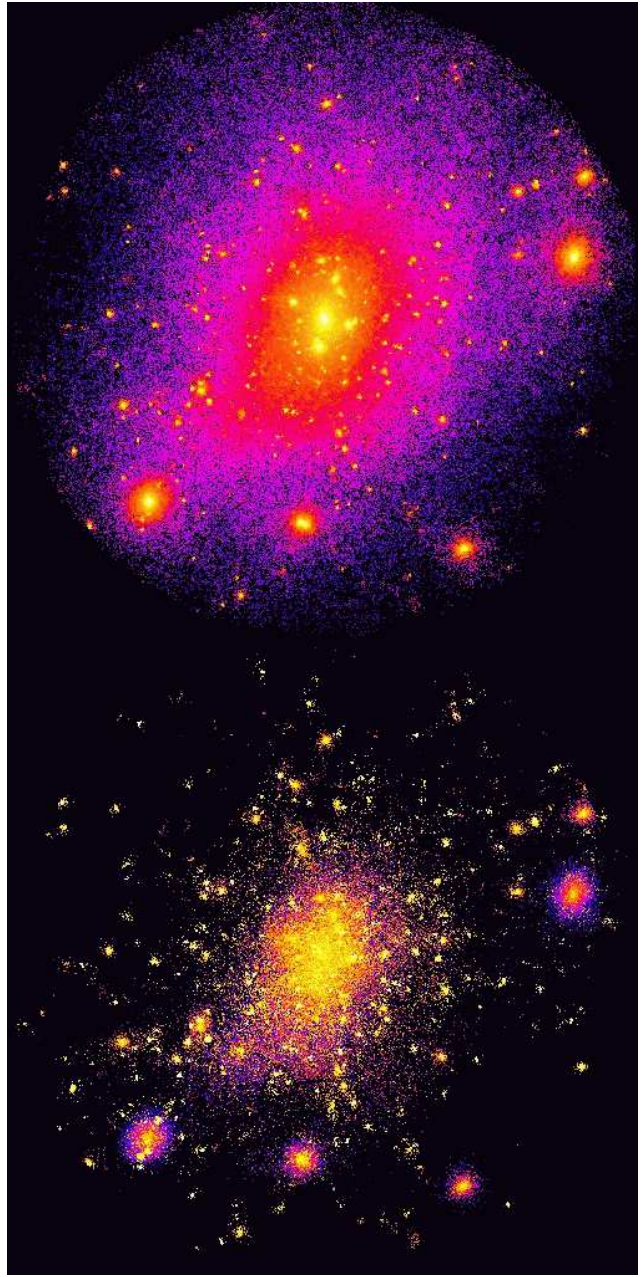}
\caption{\label{pic.ps}Maps of the cluster's 
density (top) and relaxation since virialisation (bottom) out to $r_{200}$
for the C3 run ($N_{200} \simeq 650'000$). 
The logarithmic scale for the degree of relaxation
goes form 0.01 (black) to 100 (white).}
\end{figure}

\begin{figure}
\vskip 3.2 truein
\includegraphics{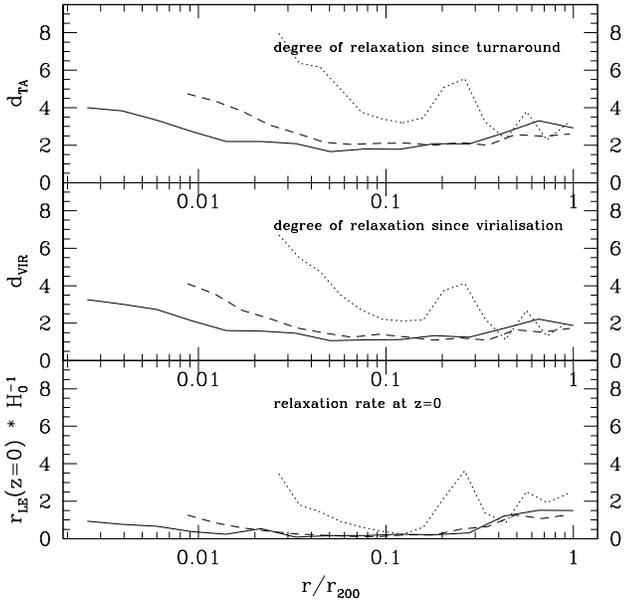}
\caption{\label{CallN.pro.ps}Relaxation versus radius. The solid line is for run C3,
the dashed line for run C2 and the dotted line for run C0.}
\end{figure}

Figure \ref{CallN.pro.ps} shows the relaxation in the cluster
as a function of the final particles position, for three 
different resolutions. In the outer part ($r \gsim 0.1 r_{200}$) the cluster
has substructure, which are small $N$ systems that exist at the present time,
so they are still relaxing at a high rate at $z=0$ (bottom panel).
Substructure haloes with $N\simeq 500$ can 
reach averages of $d_{\rm{VIR}} \simeq 10$ in all 
runs, the highest peaks in $d_{\rm{VIR}}$ are found in the centre
of substructure haloes, where $d_{\rm{VIR}}$ can be as high as $100$, 
much higher than in the centre of the host halo (see Figure \ref{pic.ps}).

Note that the degrees of relaxation (top and middle panel)
are much larger than what you would estimate using the final
distribution of particles (bottom panel).
Other studies consider only the relaxation rate at $z=0$, and claim to resolve a halo 
down to a radius where this relaxation time $r_{LE}^{-1}(z=0)$
is larger than Hubble time \citep{Power} 
or larger than three Hubble times \citep{Fukushige}.
This radius scales $\propto N^{-0.5}$, whereas convergence in
N-body simulations seem to be slower; In
\citet{Moore1998} and \citet{Ghigna} the resolved
radii are determined by comparing density profiles between simulations with
different numbers of particles.
They found that the ``resolved radius'' 
$r \simeq 0.5 ( N_{200} / V_{200} )^{-1/3}$ for a wide range of $N_{200}$
fromm $10^2$ to $10^{5.7}$. It appears like the resolved radius scales in 
the same way as the average degree of relaxation, but further relaxation studies 
for a wider range $N$ are needed to verify this.

\subsection{Mass and time dependence}

\begin{figure}
\vskip 3.2 truein
\includegraphics{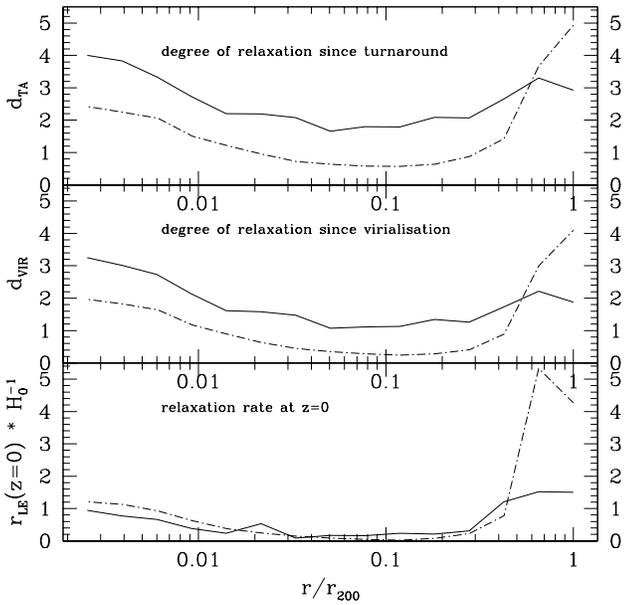}
\caption{\label{CandG.pro.ps}Relaxation vs. radius. The solid line is
the cluster run C3 ($N_{vir} \simeq 650'000$), the dot - dashed line
for the galaxy ($N_{vir} \simeq 250'000$).}
\end{figure}

Figure \ref{CandG.pro.ps} compares the relaxation rate within the high resolution
cluster and the galaxy simulations. The average degree of relaxation at $z=0$ for 
the galaxy ($\overline{d_{\rm{VIR}}} \simeq 0.58$) is much smaller than for the cluster
($\overline{d_{\rm{VIR}}} \simeq 1.34$), even though $N$ is larger for the cluster
and therefore the present relaxation rate $r_{LE}(z=0)$ is smaller in the cluster.
The reason is that the cluster forms much later than the galaxy therefore most
of its particles have spent a longer period of time in small $N$ progenitor
haloes.

\begin{figure}
\vskip 3.2 truein
\includegraphics{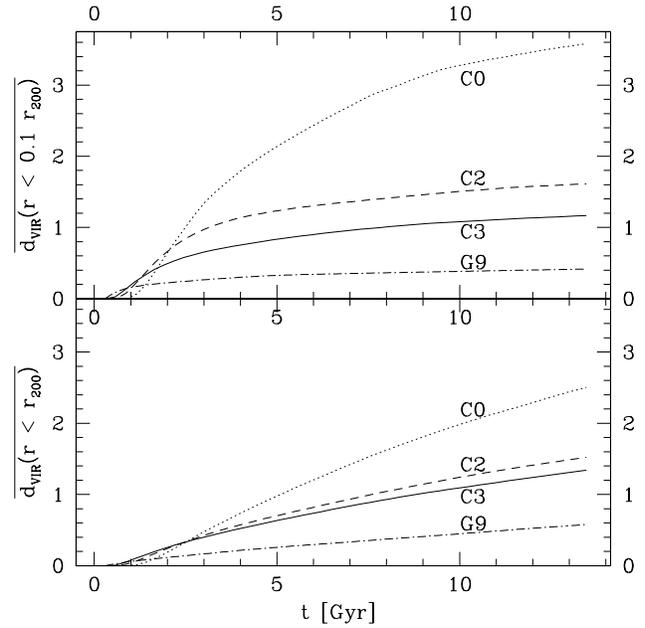}
\caption{\label{virrelVsT.ps}The degree of relaxation, $d_{\rm{VIR}}$, averaged over
all particles as a function of time. In the top panel
we average over particles inside $0.1 \,r_{200}$ at $z=0$ and in the bottom panel 
over all inside $r_{200}$.}
\end{figure}

\begin{figure}
\vskip 3.2 truein
\includegraphics{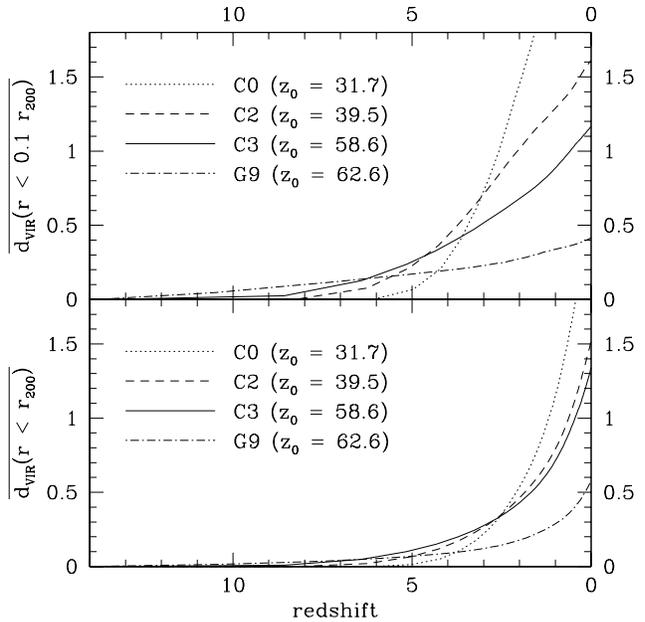}
\caption{\label{virrelVsRedshift.ps}Same as Figure \ref{virrelVsT.ps} 
but as a function of redshift. $z_0$ is the starting redshift of the runs.}
\end{figure}

In Figures \ref{virrelVsT.ps} and \ref{virrelVsRedshift.ps} 
we plot how the degree of relaxation increases with time for 
both  haloes for particles within 10 per cent of the virial radius and for all the particles within the
virial radius. 
Most of the relaxation within the central region of the galaxy occurs within the first
couple of Gyrs of the evolution of the Universe. The cluster forms over a longer timescale
and this is reflected in the longer increase in the degree of relaxation with time. 
 
The cluster runs (C0,C2,C3) 
show how relaxation in small $N$ groups starts earlier, after 1 Gyr the highest resolution 
run (C3) is most affected by relaxation. This result is not an artifact from using a
density threshold, we checked that the $d_0$ shows the same behaviour as $d_{VIR}$.
The entire haloes (bottom panel) show some relaxation during the whole simulation which arises 
from the poorly resolved substructure haloes in the outer regions.

\section{Conclusions and discussion}\label{discussion}

$N$-body cosmological simulations attempt to model a collisionless system of particles
using a technique that is inherently collisional on small scales. We have examined the
relaxation rates of isolated equilibrium cuspy haloes as a function of particle number, 
radius and softening parameter. Our results apply primarily to n-body, $P^3M$
codes, such as direct or treecodes. Adaptive grid based methods, such as
ART \citep*{Kravtsov} and MLAPM \citep*{Knebe} also seem to suffer from relaxation but 
at slightly lower rates than from those quantified here \citep{Binney2002}.

We show how one can define a local relaxation timescale
for each particle by measuring its local phase space density which we applied to 
cosmological simulations in an attempt to determine the regions most affected by numerical
relaxation. We summarise our results here:

\begin{enumerate}
\item  The relaxation rates in cuspy dark matter haloes are in good agreement 
with the rates predicted by the orbit averaged Fokker-Planck equation. However the
average relaxation time is an order of magnitude less than that measured at the half mass
radius.

\item  We verify that the average relaxation time of a halo is proportional to 
the number of particles it contains and to the inverse 
natural logarithm of the softening parameter.

\item  The relaxation time is proportional to the local phase space density which allows
us to measure the cumulative amount of relaxation each particle undergoes during the evolution
of a halo. 

\item  We show that we can measure the relaxation rate in collapsing or non-equilibrium
haloes that have kinetic to potential energy ratios up to ten times smaller
than the equilibrium value.

\item  Averaging over several simulations of cuspy Hernquist haloes we show that within 
few mean relaxation times the central cusp is transformed into a constant density core.

\item  We show that the hierarchical build up of galaxy or cluster mass haloes leads
to a greatly enhanced degree of relaxation within their central regions. The substructure
haloes suffer from the highest rates of relaxation since they contain the fewest particles
for the longest period of time. Subhaloes are typically several relaxation times old
therefore one should be cautious about interpreting their internal structure using
simulations of order $10^6$ particles \citep*{Stoehr}.

\item  Cluster haloes suffer three times the amount of relaxation as galaxy haloes simulated at
the same relative spatial and mass resolution. This is because the cluster forms later and more of its
particles spend time in poorly resolved progenitors.

\item  Increasing the resolution ($N$) at a fixed force softening reduces the
the accumulated amount of relaxation. The average degree of relaxation in CDM haloes at $z=0$
scales $\propto N^{-0.2}$, in the inner 10 percent $\propto N^{-0.3}$.
Relaxation may therefore provide a simple explanation for the slow convergence 
(resolved radius $\propto N^{-1/3}$) in density profiles of CDM haloes 
simulated at different resolutions (\citealt{Moore1998}; \citealt{Ghigna}). 

\item  Most of the affected particles become ``relaxed'' very early and within the
first few Gyrs of the evolution of the Universe. This is hardest epoch to accurately
resolve in a cosmological simulation since the relative force errors can be large and
the densities of forming haloes can be very high.
\end{enumerate}

The high degrees relaxation show that at $z=0$ many particles have completely different
energies and orbits compared to their evolution in the mean field limit ($N \to \infty$). 
Current cosmological simulations cannot model all the subtle dynamics like orbital resonances
which can be important e.g. during tidal stripping \citep{Weinberg}.
But does relaxation also affect the coarse structure of the object,
e.g. their density profiles? 
It is unclear as to how one should interpret these results for the following reason.
The highest rates of relaxation are accumulated early during the formation of the haloes.
Once a subhalo falls into a larger system the particles achieve a higher energy and it is
not clear that one should accumulate the relaxation timescale in the way that
we have done since the final ``hot'' system may lose the memory of the initial
conditions through violent relaxation processes. Indeed, \citet{Moore1999} show
that the initial conditions play little role in determine the final gross structure
of dark matter haloes.

The technique of accumulating the relaxation times by measuring the local phase
density works very well for near equilibrium structures. In some instances, the energies
of particles in a cosmological simulation will increase due to the mass increase
from merging. From the virial theorem the growth in mass must be accompanied by an increase
in velocity dispersion. Thus it is not clear how relaxation at a high redshift propagates
to the final time. However we are mainly interested in the structure of the central 
dark matter halo and the substructure haloes. These are the regions where most 
effort is going into comparing theoretical predictions with observational data
(e.g. \citealt{Moore1994}; \citealt{Stoehr}). 

The velocity dispersion in the substructure
haloes actually decreases slowly with time as mass is stripped. On the other hand
the velocity dispersion of the central halo increases with time as it grows by merging.
We can quantify the decrease in the mean velocity dispersion in subhaloes or the
central halo region by extrapolating the particles backwards
through time to examine the velocity dispersion in the progenitor haloes.
For the galaxy cluster, half of the relaxation is accumulated since $z=3$. We
therefore trace all the particles in the central few percent of the cluster
back to $z=4$ where we find that all the particles lie in the most massive
collapsed haloes at that time -- the most massive structures at high
redshift form the central cluster region through natural biasing. We find that
for the central cluster particles, $\sigma(z=4)/\sigma(z=0)=1.6$, i.e. the
mean dispersion only increases by 60\%.
Similarly for cluster substructure haloes, the ratio is practically constant through
time. For this reason we are confident that the accumulated degrees of relaxation
are a good indicator of the true relaxation of the interesting regions
within the simulations.

We note that the following thought experiment illustrates a possible way in which
relaxation in the first haloes can affect the central structure of the final halo,
even though the energies of the particles may have changed over time:
The first haloes to form are several relaxation times old and they
achieve this in an near equilibrium state. We have demonstrated that our
cumulative estimator works very well for non-equilibrium
systems that have P.E./K.E. ratios as large as ten times the equilibrium
mean (Figure 5). We also show that these first poorly resolved 
haloes will develop constant density cores through relaxation 
(Figure 8). Now consider what happens when this halo accretes 
into a larger system. Because of the constant
density core the satellite will completely disrupt at some 
distance from the centre of the final object \citep*{Moore1996}. 
If the accreting halo had a steep cusp resolved with more particles 
then it may sink deeper within the potential and 
deposit mass at smaller radii
(e.g. \citealt{Barnes}, \citealt{Syer}). Thus the early relaxation could
affect the final density profile even if the relative energies
of the particles were initially quite different from their final energies.
Unfortunately, quantifying this effect and determining if it is at
all important is best achieved by increasing the resolution by
several orders of magnitude over the haloes simulated in this paper.

\section*{Acknowledgments}

We thank Adrian Jenkins and Alexander Knebe for helpful suggestions and comments. 
We also would like to thank the Swiss supercomputing centre at Manno for computing time 
where many of these numerical simulations were performed.

\bsp
\label{lastpage}

\begin{thebibliography}{99}

\bibitem[\protect\citeauthoryear{Barnes}{1999}]{Barnes}
Barnes J. E., 1999, in Barnes J. E., Sanders D. B., eds, Proc. IAU Symp. 186,
Galaxy Interactions at Low and High Redshift. Kluwer, Dordrecht, p. 137

\bibitem[\protect\citeauthoryear{Bertin}{2000}]{Bertin}
Bertin G., 2000, Galactic Dynamics. Cambridge Univ. Press, Cambridge

\bibitem[\protect\citeauthoryear{Bertschinger}{2001}]{Bertschinger}
Bertschinger E., 2001, ApJSS, 137, 1

\bibitem[\protect\citeauthoryear{Binney \& Knebe}{2002}]{Binney2002}
Binney J., Knebe A., 2002, MNRAS, 333, 378

\bibitem[\protect\citeauthoryear{Binney \& Tremaine}{1987}]{Binney1987}
Binney J., Tremaine S., 1987, Galactic Dynamics.
Princeton Univ. Press, Princeton 

\bibitem[\protect\citeauthoryear{Chandrasekhar}{1942}]{Chandrasekhar}
Chandrasekhar S., 1942, Principles of Stellar Dynamics.
Univ. Chicago Press, Chicago 

\bibitem[\protect\citeauthoryear{Fukushige \& Makino}{2001}]{Fukushige}
Fukushige T., Makino J., 2001, ApJ, 557, 533

\bibitem[\protect\citeauthoryear{Ghigna et al.}{2000}]{Ghigna}
Ghigna S., Moore B., Governato F., Lake G., Quinn T., Stadel J., 2000, ApJ, 544, 616

\bibitem[\protect\citeauthoryear{Hayashi et al.}{2003}]{Hayashi}
Hayashi E., Navarro J. F., Taylor J. E., Stadel J., Quinn T., 
2003, ApJ, 584, 541

\bibitem[\protect\citeauthoryear{Hernquist}{1990}]{Hernquist}
Hernquist L., 1990, ApJ, 356, 359

\bibitem[\protect\citeauthoryear{Huang, Dubinski \& Carlberg}{Huang et al.}{1993}]{Huang}
Huang S., Dubinski J., Carlberg R. G., 1993, ApJ, 404, 73

\bibitem[\protect\citeauthoryear{Jing \& Suto}{2000}]{Jing}
Jing Y., Suto Y., 2000, ApJ, 529, L69

\bibitem[\protect\citeauthoryear{Kazantzidis, Magorrian \& Moore}{Kazantzidis et al.}{2003}]{Kazantzidis} 
Kazantzidis S., Magorrian J., Moore B., 2004, ApJ, 601, 37

\bibitem[\protect\citeauthoryear{Klypin et al.}{2001}]{Klypin}
Klypin A., Kravtsov A. V., Bullock J. S., Primack J. R., 2001, ApJ, 554, 903

\bibitem[\protect\citeauthoryear{Knebe, Green \& Binney}{Knebe et al.}{2001}]{Knebe}
Knebe A., Green A., Binney J., 2001, MNRAS, 325, 845

\bibitem[\protect\citeauthoryear{Kravtsov, Klypin \& Khokhlov}{Kravtsov et al.}{1997}]{Kravtsov}
Kravtsov A. V., Klypin A. A., Khokhlov A.M., 1997, ApJS, 111, 73

\bibitem[\protect\citeauthoryear{Navarro, Frenk \& White}{Navarro et al.}{1996}]{Navarro}
Navarro J. F., Frenk C. S., White S. D. M., 1996, ApJ, 462, 563

\bibitem[\protect\citeauthoryear{Moore}{1994}]{Moore1994}
Moore B., 1994, V.370, Nat, 370, 629

\bibitem[\protect\citeauthoryear{Moore, Katz \& Lake}{Moore et al.}{1996}]{Moore1996}
Moore B., Katz N., Lake G., 1996, ApJ, 457, 455

\bibitem[\protect\citeauthoryear{Moore et al.}{1998}]{Moore1998}
Moore B., Governato F., Quinn T., Stadel J., Lake G., 1998, ApJ, 499, L5

\bibitem[\protect\citeauthoryear{Moore et al.}{1999}]{Moore1999}
Moore B., Quinn T., Governato F., Stadel J., Lake G., 1999, MNRAS, 310, 1147

\bibitem[\protect\citeauthoryear{Moore et al.}{2001}]{Moore2001}
Moore B. 2001, In Martel H. \& Wheeler J., eds, AIP Conf. Proc. 586, The dark matter crisis, p.73

\bibitem[\protect\citeauthoryear{Power et al.}{2003}]{Power}
Power C., Navarro J. F., Jennkins A., Frenk C. S., White S. D. M.,
2003, MNRAS, 338, 14

\bibitem[\protect\citeauthoryear{Quinlan}{1996}]{Quinlan}
Quinlan G. D., 1996, New Astronomy, vol. 1, no. 3, 255

\bibitem[\protect\citeauthoryear{Spitzer}{1987}]{Spitzer} 
Spitzer L., 1987, Dynamical Evolution of Globular Clusters.
Princeton Series in Astrophysics, Princeton

\bibitem[\protect\citeauthoryear{Stadel}{2001}]{Stadel}
Stadel J., 2001, PhD thesis, U. Washington

\bibitem[\protect\citeauthoryear{Stoehr et. al}{2002}]{Stoehr}
Stoehr F., White S. D., Tormen G., Springel V., 2002, MNRAS, 335, L84

\bibitem[\protect\citeauthoryear{Syer \& White}{1998}]{Syer}
Syer D., White S. D. M., 1998, MNRAS, 293, 337

\bibitem[\protect\citeauthoryear{Theis}{1998}]{Theis}
Theis C., 1998, A\&A, 330, 1180

\bibitem[\protect\citeauthoryear{Weinberg}{1998}]{Weinberg}
Weinberg M. D., 1998, MNRAS, 297, 101

\end{thebibliography}
\end{document}